\documentclass[lineno]{biometrika}

\usepackage{amsmath}

\DeclareMathOperator*{\argmin}{arg\,min}
\usepackage{amsfonts}
\usepackage{hyperref}
\usepackage{caption}
\usepackage{times}
\usepackage{bm}
\usepackage{natbib}
\usepackage{xcolor}

\usepackage{newtxtext}
\usepackage[subscriptcorrection]{newtxmath}

\graphicspath{{./art/}}

\usepackage[plain,noend]{algorithm2e}

\makeatletter
\renewcommand{\algocf@captiontext}[2]{#1\algocf@typo. \AlCapFnt{}#2} 
\def\@algocf@capt@plain{top}
\renewcommand{\algocf@makecaption}[2]{%
  \addtolength{\hsize}{\algomargin}%
  \sbox\@tempboxa{\algocf@captiontext{#1}{#2}}%
  \ifdim\wd\@tempboxa >\hsize
    \hskip .5\algomargin%
    \parbox[t]{\hsize}{\algocf@captiontext{#1}{#2}}
  \else%
    \global\@minipagefalse%
    \hbox to\hsize{\box\@tempboxa}
  \fi%
  \addtolength{\hsize}{-\algomargin}%
}
\makeatother

\usepackage{xr}
\begin{document}
\nolinenumbers
\jname{Submitted to Biometrika}


\markboth{JIAZHEN XU et~al.}{Robust functional principal component analysis for non-Euclidean random objects}

\title{Robust functional principal component analysis for non-Euclidean random objects}

\author{JIAZHEN XU, ANDREW T.A. WOOD \and TAO ZOU}
\affil{Research School of Finance, Actuarial Studies and Statistics, Australian National University, Canberra ACT 2600, Australia
\email{jiazhen.xu@anu.edu.au} \hskip 0.1truein \email{andrew.wood@anu.edu.au} \hskip 0.1truein \email{tao.zou@anu.edu.au}}
\maketitle

\begin{abstract}
Functional data analysis offers a diverse toolkit of statistical methods tailored for analysing samples of real-valued random functions. Recently, samples of time-varying random objects, such as time-varying networks, have been increasingly encountered in modern data analysis. These data structures represent elements within general metric spaces that lack local or global linear structures, rendering traditional functional data analysis methods inapplicable. Moreover, the existing methodology for time-varying random objects does not work well in the presence of outlying objects. In this paper, we propose a robust method for analysing time-varying random objects. Our method employs pointwise Fr\'{e}chet medians and then constructs pointwise distance trajectories between the individual time courses and the sample Fr\'{e}chet medians. This representation effectively transforms time-varying objects into functional data. A novel robust approach to  functional principal component analysis, based on a Winsorized $U$-statistic estimator of the covariance structure, is introduced. The proposed robust analysis of these distance trajectories is able to identify key features of object trajectories over time and is useful for downstream analysis. To illustrate the efficacy of our approach, numerical studies focusing on (i) dynamic networks and (ii) time-varying spherical data are conducted. The results indicate that the proposed method exhibits good all-round performance and  surpasses the existing approach in terms of robustness, showcasing its superior performance in handling time-varying objects data.
\end{abstract}


\section{Introduction}

\subsection{Background}

In the era of big data, it is increasingly common to observe complex data over time, important examples being dynamic traffic networks, time-evolving social networks and functional Magnetic Resonance Imaging; see e.g. \cite{worsley2002general} and \cite{kolar2010estimating}.  While such time-varying data have similarities with functional data, the observations at each time point differ from the scalar or vector values typically seen in functional data analysis. Instead, they locate within a more general metric space. Unlike vector spaces, such general metric spaces lack well-defined operations such as addition, scalar multiplication, or inner products, posing a significant challenge to traditional analysis methods.

For the analysis of  non-Euclidean data, numerous works have concentrated on smooth Riemannian manifolds, typically exploiting their local Euclidean properties; see \cite{schiratti2015learning}, \cite{dai2018principal}, \cite{dai2021modeling} and \cite{shao2022intrinsic}. As these approaches predominantly focus on smooth metric spaces, they are not applicable for the analysis of data objects in more general metric spaces that do not have a natural Riemannian geometry. This limitation has sparked the development of novel and innovative approaches proposed by \cite{dubey2020functional,dubey2021modeling}. The approach considered in the latter paper, referred to below as Dubey and M{\"u}ller's method,  focuses on the squared distance of time-varying random objects from the mean trajectory. By using these  distance trajectories, Dubey and M{\"u}ller's method converts time-varying random objects into functional data, which gives access to the techniques of functional data analysis. This fruitful approach to the analysis of distance trajectories has unveiled compelling insights into the behaviour of time-varying random objects across various applications.   Nevertheless, a serious drawback of Dubey and M{\"u}ller's method is that it can be highly sensitive  to atypical curves, i.e., outlying time-varying random objects. 

For a broad and  general account of the relatively new and fast-developing field of object oriented data analysis, see the monograph  \cite{Marron2021}.

\subsection{Contributions of the paper}
Our goal in this paper is to provide a novel robust methodology for analysing time-varying random objects. Our method first constructs distance trajectories from the individual sample functions to the Fr\'{e}chet median trajectory which consists of the pointwise Fr\'{e}chet medians. These distance trajectories, which we refer to as Fr\'{e}chet median distance trajectories, are different from the Fr\'{e}chet variance trajectories in Dubey and M{\"u}ller's method, in that we measure the distance from the individual sample functions to the Fr\'echet median function, whereas the Dubey and Muller method works with the pointwise  squared distances from the individual sample functions to the sample  Fr\'echet mean function. We then develop a robust functional principal component  method  for Fr\'{e}chet median distance  trajectories of which a key ingredient is a novel robust method for estimating the relevant autocovariance operator, using a suitable Winsorized $U$-statistic.

We briefly explain how our proposed  procedure for robust autocovariance operator estimation goes beyond existing  approaches. Robust principal component analysis approaches for Euclidean data using spatial sign covariance and spherical principal component approaches for functional data have been studied in \cite{marden1999some}, \cite{visuri2000sign} and \cite{gervini2008robust}. However, these approaches require that data are symmetrically distributed and, in addition,  there is  scope for improving their numerical performance. \cite{taskinen2012robustifying}, \cite{han2018eca},  \cite{zhong2022robust} and \cite{wang2023robust} utilize pairwise covariance to avoid the need for symmetry requirements but these methods can be extended by introducing quantile-based weights in $U$-statistic covariance estimators, compared to spherical principal component approaches which projects the centred data onto a unit sphere. \cite{raymaekers2019generalized} and \cite{leyder2023generalized} use quantile-based weights to extend spatial sign covariance to achieve better numerical performance with respect to efficiency and estimation accuracy in Euclidean cases but they are still reliant on a symmetry assumption. These works motivate us to consider an extension of the  pairwise spatial sign autocovariance proposed in \cite{zhong2022robust} and \cite{wang2023robust}. Our approach has the potential to inspire novel types of robust principal component analysis in Euclidean and Hilbert spaces and more general object data settings.

There exist four major theoretical obstacles in working with Fr\'{e}chet median distance trajectories which, at the outset, make it challenging to directly apply functional data analysis and the existing theoretical techniques used in time-varying random objects. These are: (i) distance trajectories are non-negative, which suggests that a symmetry assumption will not be reasonable; (ii)  distance trajectories involve the unknown population Fr\'{e}chet median which needs to be estimated from data; (iii) the widely used total boundedness metric space assumption in recent literature in the present setting may be considered restrictive in robust analysis; and (iv) the asymptotic equicontinuity of sample Fr\'{e}chet median trajectories is hard to verify. We briefly explain how we have dealt with these four challenges; further details are given in the theoretical results below.

Regarding (i), various robust methodologies have been proposed  such as the spherical principal component approach of \cite{locantore1999robust} and \cite{gervini2008robust} and the projection-pursuit approach of \cite{bali2009principal}. However, most of the robust functional principal component  methods except the one in \cite{wang2023robust} are reliant on a symmetry assumption, which does not align with the asymmetric nature of the non-negative Fr\'{e}chet median distance trajectories. As already mentioned, inspired by \cite{wang2023robust}, our approach avoids the need for symmetry requirements by working with $U$-statistic-type autocovariance estimators. 

Regarding (ii), Fr\'{e}chet median distance trajectories involve the unknown population Fr\'{e}chet median trajectory and a sample version of the Fr\'{e}chet median trajectory needs to be estimated from the data. The distance trajectories which are used for the analysis are then the distances of the individual time courses from the sample Fr\'{e}chet median trajectory. The autocovariance operator for these empirical distance trajectories becomes a stochastic process with estimated parameters. Classical techniques for handling the asymptotic behaviour of stochastic processes with estimated parameters rely on Taylor series expansions (see \cite{randles1982asymptotic, arcones1994u}). However, for general metric spaces lacking local or global linear structures, Taylor series-based methods are, typically, not applicable. To study the asymptotic behaviour of the autocovariance operator, we employ techniques from empirical process theory, making use of Donsker preservation properties; see Theorem 2.10.6 in \cite{vdVW96}. By introducing a Winsorized $U$-statistic into the pairwise autocovariance operator, we are able to leverage Donsker preservation and thereby establish desirable asymptotic properties for the eigenfunction estimates, including rates of convergence.

Regarding (iii), in several recent papers, such as \cite{petersen2019frechet, dubey2021modeling, chen2022uniform}, the metric space is assumed to be bounded, while robust analysis addresses atypical observations that are possibly far away from the uncontaminated observations. Thus the global total boundedness metric space assumption is typically too restrictive when considering robust analysis. We relax this global total boundedness assumption and just assume a local form of total boundedness plus a certain moment condition, specified in Assumption \ref{condition c0} in the Appendix.

Regarding (iv), most existing works that focus on the Fr\'{e}chet mean trajectory (see \citealt{petersen2019frechet, dubey2021modeling, chen2022uniform}) establish the uniform convergence of the sample Fr\'{e}chet mean trajectory by employing empirical process techniques, where the key challenge is demonstrating the asymptotic equicontinuity of the sample Fr\'{e}chet mean. However, establishing the asymptotic equicontinuity for the sample Fr\'{e}chet median trajectories is more difficult. To overcome this, we introduce a novel methodology that ensures uniform convergence by assuming local convexity of the empirical cost function for the Fr\'{e}chet median; see detailed discussion in \S \ref{subsect::estimation} and Assumption \ref{condition c1.5} in the Appendix. Assumption \ref{condition c1.5} makes use of the idea of a convex metric space (\cite{abdelhakim2016convexity}; \cite{khamsi2011introduction}) which, so far as we are aware, is a concept new to the statistics literature.

Numerical studies in this paper showcase that our approach to functional principal components analysis using Fr\'{e}chet median distance trajectories produces robust estimation of eigenfunctions while Dubey and M{\"u}ller's method, which was not designed to be robust,  is sensitive to outliers. 
The pointwise Fr\'{e}chet medians provide a robust representative for the most central point for a sample object functions and the Fr\'{e}chet median distance trajectories carry information about the deviations of individual trajectories from the central trajectory. As demonstrated in the numerical studies, the proposed robust method is useful for cluster analysis of time-varying object data and also for the  detection of outlying objects.

\section{Preliminaries and Methodologies}\label{sect::methodology}

Consider an object space $(\Omega,d)$ that is a separable metric space such that each $\omega \in \Omega$ has a totally bounded neighbourhood, where total boundedness is defined e.g. in \cite{vdVW96}. We also consider an $\Omega$-valued stochastic process $\{X(t)\}_{t\in[0,1]}$. For uncontaminated data, assume that we observe a sample of independently and identically distributed (IID) copies of the random process $X(t)$, denoted by $X_1(t),\ldots,X_n(t)$. Motivated by Dubey and M{\"u}ller's method, for each trajectory $X_i(t)$, we focus on the measurement of its deviation from a baseline object function which, for robustness purposes, is chosen to be the Fr\'{e}chet median function. The population and sample Fr\'{e}chet median trajectories at each time point $t\in[0,1]$ are defined as
\begin{equation}\label{def::frechet median}
\mu_{\rm GM}(t)=\argmin_{\omega\in\Omega} E  [d(X(t),\omega)],~\hat{\mu}_{\rm GM}(t)=\argmin_{\omega\in\Omega}\frac{1}{n}\sum_{i=1}^n d(X_i(t),\omega),
\end{equation}
respectively. We also assume the existence and the uniqueness of these minimizers for all $t\in[0,1]$. Related discussions can be found in \cite{sturm2003probability} and \cite{ahidar2020convergence}.
Our target functions for downstream analysis will be 
\[
V_i(t)=d(X_i(t),\mu_{\rm GM}(t)),t\in[0,1],
\]
which correspond to the pointwise distance functions of the subject trajectories $X_i$ from the population Fr\'{e}chet median trajectories. Our target functions effectively transform time-varying random objects into functional data, and functional principal component analysis is standard dimensionality reduction technique in functional data analysis. It operates by utilizing the eigenfunctions of the autocovariance operator of the observed data, allowing for an efficient representation of the underlying structure in functional data. The population Fr\'{e}chet covariance function $C$ for a typical $V(t)$ is
\begin{equation}\label{orig cov}
C(s,t)= E  [V(s)V(t)]- E  [V(s)] E  [V(t)],
\end{equation}
where $V(t)=d(X(t),\mu_{\rm GM}(t)),t\in[0,1]$ is the population Fr\'{e}chet median distance trajectories.

For uncontaminated data, the eigenvalues of the autocovariance operator are non-negative as the covariance surface is symmetric and non-negative definite. By Mercer's theorem (\citealt{hsing2015theoretical}),
\begin{equation}
C(s,t)=\sum_{j=1}^\infty \lambda_j(C)\phi_j(s)\phi_j(t), s,t\in[0,1]
\label{late_addition}
\end{equation}
with uniform convergence, where $\lambda_j(C)$ denotes the $j$-th eigenvalues of the covariance operator $C$, ordered in decreasing order, and $\phi_j(\cdot)$ is the corresponding orthonormal eigenfunction. Based on the decomposition of the autocovariance operator, the Karhunen-Lo\`{e}ve  expansion of the Fr\'{e}chet median distance trajectories
\[
V(t)=\nu(t)+\sum_{j=1}^\infty B_{j}\lambda_j^{1/2}(C)\phi_j(t),
\]
holds with $L^2$ convergence, where $\nu$ is the mean function of the subject-wise Fr\'{e}chet median distance functions, the random variable $B_{j}$ is uncorrelated with $ E  (B_{j})=0$ and ${\rm var}(B_{ij})=1$. Note that $B_{j}\lambda_j^{1/2}(C)=\int (V(t)-\nu(t))\phi_j(t)dt$ are the functional principal components.

For outlying time-varying random objects $X^{\rm out}(t)$, let $M(t)=d(X(t),X^{\rm out}(t))$ be the pointwise measurement of the distance between the uncontaminated data and the outliers. One can see that the outliers $X^{\rm out}(t)$ are the ones with $\|M(\cdot)\|$  being unusually large, where $\|\cdot\|$ is the Hilbert norm. Moreover, the outlying random objects will also result in $\|D(\cdot)\|$ being unusually large where $D(t)=V(t)-V^{\rm out}(t)$ and $V^{\rm out}(t)=d(X^{\rm out}(t),\mu_{\rm GM}(t))$. To see this, note that $\mu_{\rm GM}$, the pointwise median trajectory, represents the centre of the uncontaminated time-varying random objects. Therefore, the existence of outlying objects may result in some  Fr\'{e}chet median distance trajectories not being square integrable, and the population Fr\'{e}chet covariance function $C$ given in (\ref{orig cov}) may not exist and be finite for all $s$ and $t$. This motivates us to consider an autocovariance operator using a Winsorized pairwise $U$-statistic to achieve robustness,, which is given by
\begin{equation}\label{swPCA cov}
C_{\rm WPU}(s,t)= E  \left[  \xi^2(\|V(\cdot)-\widetilde{V}(\cdot)\|)\{V(s)-\widetilde{V}(s)\}\{V(t)-\widetilde{V}(t)\} \right],
\end{equation}
where $\xi(\cdot):[0,\infty) \to (0,\infty)$ is a Winsorized radius function given by
\begin{align*}
\xi(r)=\begin{cases}
1,&r\leq Q\\
Q/r &r>Q,
\end{cases}
\end{align*}
with $Q$ being a cutoff point depending on $\|V(\cdot)-\widetilde{V}(\cdot)\|$, and $\widetilde{V}(\cdot)$ is a independent copy of $V(\cdot)$. Note that when $Q=0$, the proposed autocovariance operator (\ref{swPCA cov}) is the one combining the pairwise spatial sign operator proposed in \cite{wang2023robust} with the Fr\'{e}chet median distance trajectories, referred to below as Wang's method. However, to achieve the desirable asymptotic properties of the robust functional principal component analysis eigenfunction estimation, we require that the cutoff point $Q$ being positive and thus the radius function $\xi(\cdot)$ is bounded. By utilizing the Winsorized pairwise $U$-statistic, the autocovariance operator (\ref{swPCA cov}) will have the same set of eigenfunctions with the same order of eigenvalues as those of the regular covariance function (\ref{orig cov}), which is discussed in Corollary \ref{thm::eigendecomp}. For convenience purposes, we let $\phi_j$ be the  eigenfunction of $C_{\rm WPU}$, corresponding to the $j$-th eigenvalues.

\section{Theoretical Properties}\label{sect::theoretical properties}

\subsection{Preliminaries}  We now develop theoretical frameworks of estimation theory and analysis of robustness. In \S\ref{subsect::estimation}, we establish asymptotic properties of the empirical estimators of the population targets as described in \S\ref{sect::methodology}. In \S\ref{sub sect::robust}, we derive theoretical results of robustness, including the influence functions of eigenfunctions and the upper breakdown point of the proposed  robust autocovariance estimator. Assumptions needed for our estimation theory and analysis of robustness are presented and discussed in the Appendix. All proofs in this article are given in the supplementary material.

\subsection{Estimation}\label{subsect::estimation}

Before studying the estimation theory of the proposed Winsorized autocovariance function $C_{\rm WPU}$, we first need to establish the uniform convergence of the sample Fr\'{e}chet median $\hat{\mu}_{\rm GM}$. Let $\bar{M}_n(\omega, t)=\frac{1}{n}\sum_{i=1}^n d(\omega,X_i(t))$ be the empirical cost function of the sample Fr\'{e}chet median, Proposition \ref{prop::uniform bound from loss to estimator} provides a connection between the uniform bounds of $\ell_n$ and $d(\hat{\mu}_{\rm GM}(t),\mu_{\rm GM}(t))$.

\begin{proposition}\label{prop::uniform bound from loss to estimator}
Under Assumption \ref{condition c1.5}, for any fixed $t$, if $ \bar{M}_n(\omega, t)  - \bar{M}_n(\mu_{\rm GM}(t), t)>0$ for all $\omega\in B_\delta(\mu_{\rm GM}(t))$, where $B_\delta(\mu_{\rm GM}(t))=\{\omega\in\Omega:d(\omega,\mu_{\rm GM}(t))<\delta\}$, then $d(\hat{\mu}_{\rm GM}(t),\mu_{\rm GM}(t))<\delta$.
\end{proposition}

Proposition \ref{prop::uniform bound from loss to estimator}, which extends Lemma 9.21 in \cite{wainwright2019high}  to a class of non-Euclidean spaces, provides a powerful technique for deriving uniform convergence of $d(\hat{\mu}_{\rm GM}(t),\mu_{\rm GM}(t))$.  This is significant because it is typically difficult to prove such results directly using empirical process theory. Detailed discussion can be found in \S \ref{sect::supp discussion on uniform convergence} in the supplementary material.

Theorem \ref{prop::uniform bound to uniform convergence} below provides sufficient conditions to derive the uniform convergence of the sample Fr\'{e}chet median by utilizing Proposition \ref{prop::uniform bound from loss to estimator}. It is worth noting that both Proposition \ref{prop::uniform bound from loss to estimator} and Theorem \ref{prop::uniform bound to uniform convergence} work for a general M-estimator.

\begin{theorem}\label{prop::uniform bound to uniform convergence}
Under Assumption \ref{condition c1.5}, if for any $\delta>0$,
\begin{align}\label{improved sufficient condition}
pr\left (\inf_{t \in [0,1] } \inf_{\omega \in B_\delta(\mu_{\rm GM}(t))} \bar{M}_n(\omega, t)  - \bar{M}_n(\mu_{\rm GM}(t), t) >0 \right ) \to 1,
\end{align}
as $n \to \infty$, we have $\sup_{t \in [0,1]} d( \hat{\mu}_{\rm GM}(t) , \mu_{\rm GM}(t)) =o_p(1)$.
\end{theorem}

To establish the uniform convergence of the Fr\'{e}chet median trajectory given in Theorem \ref{thm::uniform convergence of median} below, we only need to verify the condition (\ref{improved sufficient condition}) in Theorem \ref{prop::uniform bound to uniform convergence}.

\begin{theorem}\label{thm::uniform convergence of median}
Under Assumption \ref{condition c0}-\ref{condition c1.5}, $\sup_{t\in[0,1]}d(\hat{\mu}_{\rm GM}(t),\mu_{\rm GM}(t)) =o_p(1)$.
\end{theorem}

Having  established the uniform convergence of the Fr\'{e}chet median trajectory, we are able to derive the estimation theory of the Winsorized autocovariance surface $C_{\rm WPU}$. If the population Fr\'{e}chet median trajectory $\mu_{\rm GM}$ is known, the oracle estimator of $C_{\rm WPU}$ is 
\begin{equation}\label{autocovariance oracle estimator}
\widetilde{C}_{\rm WPU}(s,t)=\frac{2}{n(n-1)}\sum_{1\leq j<k\leq n}\xi^2(d_{jk}) \{V_j(s)-V_k(s)\}\{V_j(t)-V_k(t)\}, 
\end{equation}
where $d_{jk}= \|V_j(\cdot)-V_k(\cdot)\|$. Under Assumption \ref{condition c3} in the Appendix, standard asymptotic theory from the U-process (\citealt{arcones1993limit}) that this estimator has desirable asymptotic properties and converges to the true Winsorized autocovariance surface $C_{\rm WPU}$.

However, the population Fr\'{e}chet median trajectory $\mu_{\rm GM}$ is unknown in practice, we need to use the sample Fr\'{e}chet median $\hat{\mu}_{\rm GM}$ to replace it in (\ref{autocovariance oracle estimator}). Define $\hat{d}_{jk}= \|\hat{V}_j(\cdot)-\hat{V}_k(\cdot)\|$, the sample version of $C_{\rm WPU}$ is
\[
\widehat{C}_{\rm WPU}(s,t)=\frac{2}{n(n-1)}\sum_{1\leq j<k\leq n}\xi^2(\hat{d}_{jk}) \{\hat{V}_j(s)-\hat{V}_k(s)\}\{\hat{V}_j(t)-\hat{V}_k(t)\}. 
\]
A key step to derive the limiting behaviour of $\widehat{C}_{\rm WPU}$ given in Theorem \ref{thm::asympt of sw cov} below is to show that $\widehat{C}_{\rm WPU}$ is asymptotically close to the oracle estimator $\widetilde{C}_{\rm WPU}$.

\begin{theorem}\label{thm::asympt of sw cov}
For a function $\check{\omega}\in\mathcal{H}$, let $V_{\check{\omega}}(s)=d(\omega(s),X(s))$ and let $\widetilde{V}_{\check{\omega}}(s)=d(\omega(s),\widetilde{X}(s))$ for $s\in[0,1]$, define $\theta(\check{\omega},s,t)= E  [f_{\check{\omega},s,t}(X(\cdot),\widetilde{X}(\cdot))]$, where $\widetilde{X}$ is an IID copy of $X$, and
\begin{align*}
f_{\check{\omega},s,t}(X(\cdot),\widetilde{X}(\cdot))=&\left[V_{\check{\omega}}(s)-\widetilde{V}_{\check{\omega}}(s)\right]\left[V_{\check{\omega}}(t)-\widetilde{V}_{\check{\omega}}(t)\right]\xi^2\left(\| V_{\check{\omega}}(\cdot)-\widetilde{V}_{\check{\omega}}(\cdot))\|\right).
\end{align*}
Under Assumptions \ref{condition c0}-\ref{condition c4} in the Appendix, the sequence of stochastic processes 
\[
U_n(s,t)=\sqrt{n}\left(  \hat{C}_{\rm WPU}(s,t)-C_{\rm WPU}(s,t) - \theta(\hat{\mu}_{GM},s,t) +  \theta(\mu_{GM},s,t) \right)
\]
converges weakly to a Gaussian process with mean zero and covariance function
\[
R_{(s_1,t_1),(s_2,t_2)}={\rm cov}\left( f_{s_1,t_1}(X(\cdot),\widetilde{X}(\cdot)),f_{s_2,t_2}(X(\cdot),\widetilde{X}(\cdot)) \right),
\]
where 
\begin{align*}
f_{s,t}(X(\cdot),\widetilde{X}(\cdot))=&\left[V(s)-\widetilde{V}(s)\right]\left[V(t)-\widetilde{V}(t)\right]\xi^2\left(\| V(\cdot)-\widetilde{V}(\cdot)\|\right).
\end{align*}
\end{theorem}

\begin{remark}
Theorem \ref{thm::asympt of sw cov} studies the asymptotic behaviour of stochastic processes with estimated parameters. The results align with corresponding results in finite-dimensional settings such as \cite{randles1982asymptotic,arcones1994u}. Instead of using Taylor series expansions, which may not be applicable in this case, we apply techniques coming from empirical process theory to deal with the asymptotic behaviour. It is worth noting that our techniques still apply in the 
finite-dimensional setting. Moreover, the differential of $\theta(\check{\omega},s,t)$  at $\check{\omega}=\mu_{\rm GM}$, if it exists,  is not necessarily zero, the estimator $\hat{\mu}_{\rm GM}$ will play a role in the limiting Gaussian process.
\end{remark}

\begin{remark}
    Note that the sequence of stochastic processes $U_n(s,t)$ can be split as $U_n(s,t)=I+II$ where $I=\sqrt{n}\Big(\hat{C}_{\rm WPU}(s,t)-\widetilde{C}_{\rm WPU}(s,t)- \theta(\hat{\mu}_{GM},s,t) + \theta(\mu_{GM},s,t)\Big)$ and $II=\sqrt{n}\Big(\widetilde{C}_{\rm WPU}(s,t)-C_{\rm WPU}(s,t)\Big)$. Theorem \ref{thm::asympt of sw cov} is established by showing that the first term $I$ converges to zero in probability while the second term $II$ converges weakly to a Gaussian process. Moreover, the convergence in probability of the first term aligns with Theorem 2.8 in \cite{randles1982asymptotic} in a smooth setting.
\end{remark}

Theorem \ref{thm::asympt of sw cov} shows that $ \theta(\hat{\mu}_{GM},s,t) - \theta(\mu_{GM},s,t)$ actually affects the convergence rate of the proposed estimator. The asymptotic behaviour of $ \theta(\hat{\mu}_{GM},s,t) - \theta(\mu_{GM},s,t)$ is given in our next result, which determines the uniform convergence and rates of convergence of $\sup_{t\in[0,1]}\left| \hat{\phi}_j(t)-\phi_j(t) \right|$, where $\hat{\phi}_j(t)$ is the $j$-th empirical eigenfunctions of the $\widehat{C}_{\rm WPU}$.     

\begin{theorem}\label{thm::convergence of eigenfunction}
Under Assumptions \ref{condition c1}-\ref{condition c5} in the Appendix with $\delta_j$ defined in Assumption \ref{condition c5}, we have
\[
\sup_{s,t\in[0,1]}\left| \theta(\hat{\mu}_{GM},s,t) - \theta(\mu_{GM},s,t)\right|=O_p\left( n^{-1/2} \right),
\]
and
\[
\sup_{t\in[0,1]}\left| \hat{\phi}_j(t)-\phi_j(t) \right|=O_p\left( \delta_j^{-1} n^{-1/2}  \right),
\]
if $\beta=2$ with $\beta$ defined in Assumption \ref{condition c2} in the Appendix. Otherwise, 
\[
\sup_{s,t\in[0,1]}\left| \theta(\hat{\mu}_{GM},s,t) - \theta(\mu_{GM},s,t)\right|=O_p\left( \max\left\{n^{-1/2} , \left( n^{-2} \log (n) \right)^{1/\{4(\beta-1)\}} \right\}\right),
\]
and
\[
\sup_{t\in[0,1]}\left| \hat{\phi}_j(t)-\phi_j(t) \right|=O_p\left( \delta_j^{-1} \max\left\{n^{-1/2}  , \left( n^{-2} \log (n) \right)^{1/\{4(\beta-1)\}} \right\}\right).
\]
\end{theorem}

It is worth noting that object spaces that satisfy Assumption \ref{condition c2} in the Appendix with $\beta=2$ include graph Laplacians of networks with the Frobenius metric, univariate probability distributions with the 2-Wasserstein metric and correlation matrices of a fixed dimension with the Frobenius metric (\citealt{petersen2019frechet,dubey2021modeling}). 

\subsection{Analysis of Robustness}\label{sub sect::robust}

The first result in this subsection shows that, the Winsorized autocovariance operator (\ref{swPCA cov}) has the same set of eigenfunctions as those of the regular autocovariance function (\ref{orig cov}). In addition, although the eigenvalues of the Winsorized autocovariance operator are generally different from those of the regular autocovariance function, they remain of the same order under an additional distributional assumption.

\begin{corollary}\label{thm::eigendecomp}
Under Assumption \ref{condition c6} in the Appendix, the Winsorized autocovariance function (\ref{swPCA cov}) admits the following decomposition
\begin{equation}\label{eq::decomp of SW autocoviance}
C_{\rm WPU}(s,t)=\sum_{j=1}^\infty \lambda_j(C_{\rm WPU})\phi_j(s)\phi_j(t), s,t\in[0,1],
\end{equation}
where $\phi_j$ is the $j$-th eigenfunction of the regular autocovariance function given in (\ref{late_addition}), for any $j$ such that $\lambda_j(C)>0$,
\[
\lambda_j(C_{\rm WPU})=\lambda_j(C) E  \left[  Z_j^2\xi\left(\sum_{l=1}^\infty Z_l^2\lambda_l(C) \right)  \right],
\]
with $Z_j = \langle V - \widetilde{V}, \phi_j\rangle / \lambda_j^{1/2}(C)$ and $\langle\cdot,\cdot\rangle$ is the inner product operator in the relevant Hilbert space.   If $Z_j$, $j=1,2,\ldots$, are further assumed to be exchangeable, then for any positive integers $j,k$, $\lambda_{j}(C)\leq\lambda_k(C)$ implies that $\lambda_j(C_{\rm WPU})\leq \lambda_k(C_{\rm WPU})$.
\end{corollary}
\begin{remark}
Corollary \ref{thm::eigendecomp} provides analogous findings to those in \cite{wang2023robust}. Corollary \ref{thm::eigendecomp} is established for the autocovariance operator using the Winsorized $U$-statistics, which can be viewed as an extension of the pairwise spatial sign operator in \cite{wang2023robust}.
\end{remark}

Decomposition (\ref{eq::decomp of SW autocoviance}) in Corollary \ref{thm::eigendecomp} provides the theoretical basis for using the sample version of $C_{\rm WPU}$ as a robust estimator of the eigenfunctions  of $C$ defined in (\ref{orig cov}) and (\ref{late_addition}).

One of the measures of robustness is the influence function (\citealt{huber2004robust}), which measures the sensitivity of an estimator to clustered outliers. Let $\Phi_k(P)$ be the $k$-th principal component of a random process with distribution $P$. The influence function is defined as ${\rm IF}_{\Phi_k}(z(\cdot))=\lim_{\varepsilon\to 0} \varepsilon^{-1}\{\Phi_k(P_{\varepsilon,z(\cdot)})-\Phi_k(P_0)\}$, where $P_{\varepsilon,z(\cdot)}=(1-\varepsilon)P_0+\varepsilon\delta_{z(\cdot)}$ and $\delta_{z(\cdot)}$ is the point-mass probability at $z(\cdot)\in L^2([0,1])$. The next theorem gives the influence function of $\Phi_k$.

\begin{theorem}\label{thm:: IF}
Suppose that Assumption \ref{condition c6} holds. When $z(\cdot)=\mu_{\rm GM}$,
\[
{\rm IF}_{\Phi_k}(z(\cdot))=0.
\]
When $z(\cdot)\neq \mu_{\rm GM}$,
\begin{align*}
{\rm IF}_{\Phi_k}(z(\cdot))=2\sum_{j\neq k}  \left (\lambda_k(C_{\rm WPU}) - \lambda_j(C_{\rm WPU}) \right )^{-1}  E  _{P_0}[\zeta_j(z(\cdot))\zeta_k(z(\cdot))]\phi_j ,
\end{align*} 
where 
\[
\zeta_j(z(\cdot))=\left\langle d(X(\cdot),\mu_{\rm GM})-d(z(\cdot),\mu_{\rm GM}) , \phi_j  \right\rangle \xi\left( \|  d(X(\cdot),\mu_{\rm GM})-d(z(\cdot),\mu_{\rm GM})   \| \right).
\]
Moreover, the gross-error sensitivity is 
\[
\gamma_{\Phi_k}^*=\sup \left\{ \|{\rm IF}_{\Phi_k}(z(\cdot))\|:z(\cdot)\in L^2([0,1]) \right\}=\{p_1+Q(1-p_1)\}/c_k,
\]
where $c_k=\min \{ \lambda_k(C_{\rm WPU}), \{ \min |\lambda_k(C_{\rm WPU})-\lambda_j(C_{\rm WPU})|:j=1,\ldots,p, j\neq k \} \}$ and $p_1=pr\Big( \|  d(X(\cdot),\mu_{\rm GM})-d(z(\cdot),\mu_{\rm GM})   \|>Q\Big)$.
\end{theorem}
This theorem indicates the influence function of $\Phi_k$ depends strongly on the eigenvalue spacings. If we have a sequence of distinct eigenvalues that means Assumption \ref{condition c5} in the Appendix holds, the principal components will be robust as the gross-error sensitivity is bounded.

Note that the autocovariance operator $C_{\rm WPU}$ involves a cutoff point $Q$ and the choice of $Q$ will affect the breakdown point of $C_{\rm WPU}$. Define $\Delta=\|V(\cdot)-\widetilde{V}(\cdot)\|$. The upper breakdown of $C_{\rm WPU}$ operator is given by the following theorem.
\begin{theorem}\label{thm::UB autocovariance}
Suppose that $\Delta=\|V(\cdot)-\widetilde{V}(\cdot)\|$ has a cumulative distribution function $G$ with $\alpha$-quantile $Q_\alpha$, i.e., $Q_\alpha=\min\{ q:pr(\Delta\leq q)\geq \alpha \}$. If we choose the cutoff point $Q=Q_\psi$, then the upper breakdown of $C_{\rm WPU}$ is $(1-\psi)^{1/2}$. That is, $n(1-\psi)^{1/2}$ is the smallest number of bad observations that can force $C_{\rm WPU}$ to breakdown, in the sense of the order of eigenvalues of $C_{\rm WPU}$.
\end{theorem}

In practice, we do not know $Q_\psi$, but it can be consistently estimated as follows in finite samples. Given a sample $\hat{V}_1(\cdot),\ldots,\hat{V}_n(\cdot)$, define 
\[
\hat{\Delta}_{ij}=\|\hat{V}_i(\cdot)-\hat{V}_j(\cdot)\|,~1\leq i < j \leq n.
\]
Define $N=n(n-1)/2$ and write
\[
\hat{\Delta}_{(1)}\leq \hat{\Delta}_{(2)}\leq\ldots\leq \hat{\Delta}_{(N)}
\]
for the ordered values of $\{\hat{\Delta}_{ij}\}$. Then, the following corollary shows that, $\hat{Q}$, a consistent estimator of $Q_\psi$, is given by $\hat{Q}=\hat{\Delta}_{(m(\psi))}$, where $m(\psi)=n\psi(n-1)/2$.

\begin{corollary}\label{corollary::quantile consistency}
Under Assumptions \ref{condition c0}-\ref{condition c4} in the Appendix, $\hat{Q}$ convergences in probability to $Q_\psi$ as $n\to\infty$.
\end{corollary}

\section{Case Study}\label{sect::case study}

The New York City Citi Bike sharing system provides historical bike trip data publicly, accessible at \url{https://citibikenyc.com/system-data}. This data set records the start and end times of trips, along with start and end locations, at a second-resolution level, encompassing trips between bike stations in New York City. 

In this study, we analyse trip records from January 2017 through June 2020, focusing on the day-to-day dynamics of bike rides between stations to better understand transportation patterns and the operational characteristics of the Citi Bike system. We've selected the top 90 popular stations and divided each day into 20-minute intervals. Within each 20 minute interval, a network is constructed comprising 90 nodes (corresponding to the selected bike stations). Edge weights represent the number of recorded bike trips between the pairs of stations, forming the time-varying network for each of the 1273 observed days in the years 2017 to 2020. The network is sampled at the midpoint of each 20-minute interval, and at each time point, the network structure is represented by a 90-dimensional graph Laplacian, describing the number of trips between the 90 stations during that interval. The graph Laplacians $L$ for a network with $p$ nodes is obtained via $L = D-A$, where $A$ is the $p\times p$ adjacency matrix with the $(i,j)$-th entry $a_{ij}$ representing the edge weight between nodes $i$ and $j$, and $D$ is the degree matrix, the off-diagonal entries of which are zero, with diagonal entries $d_{ii}=\sum_{j=1}^p a_{ij}$. The graph Laplacian determines the simple network uniquely.

\begin{figure}
\centering
\includegraphics[scale=0.55]{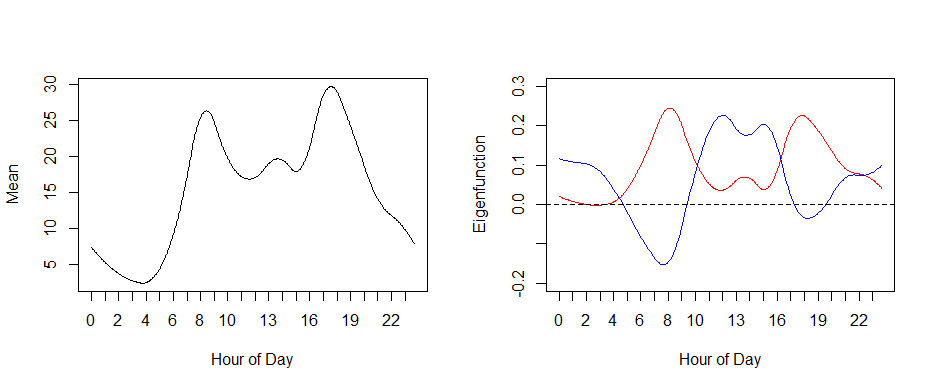}
\caption{Sample mean function (left plot) and eigenfunctions for the robust functional component analysis (right plot) of the  distance trajectories at 20-minute intervals of graph Laplacians of daily Citi Bike trip networks in New York City. In the right plot, the red line represents the first eigenfunction, which  accounts for 59.25\% of variability in the trajectories and the blue line represents the second eigenfunction, explaining 13.54\% of the variability.\label{fig::sample mean and first second eigenfunction}}
\end{figure}

In the present case study, the distance between graph Laplacians is calculated using the Frobenious norm, an extrinsic metric. The sample Fr\'{e}chet median at each time point is chosen to minimize the relevant sum of Frobenious  distances. We then obtained the daily Fr\'{e}chet median distance trajectories, whose value at each time point reflects the Frobenius distance between the graph Laplacian and the Fr\'{e}chet median graph Laplacian. The resulting robust estimator of the autocovariance function was applied to these 1273 Fr\'{e}chet median distance trajectories. The mean Fr\'{e}chet median distance trajectory of the daily graph Laplacians for the Citi Bike trip networks as a function of the time within the day, which captures the average deviation from the median trajectory, is shown in the left plot of Fig. \ref{fig::sample mean and first second eigenfunction}. Notable peaks appear at 8 am, with elevated mean variation  from 6 am to 10 am, and again at 6 pm, with higher levels between 4 pm and 7 pm, which reflect the morning and late afternoon commuting surges, when network variability is at its highest.

The first two eigenfunctions of  the robust autocovariance function estimator, elucidated in the right plot of Fig. \ref{fig::sample mean and first second eigenfunction}, explain about 72.79\% of the variation in the trajectories. The first eigenfunction reflects increased variability around the peaks of the Fr\'{e}chet median  function that is shown in Fig. \ref{fig::sample mean and first second eigenfunction}, aligning with commuter rush hours.

\begin{figure}
\begin{minipage}[c]{0.33\linewidth}
\includegraphics[width=\linewidth]{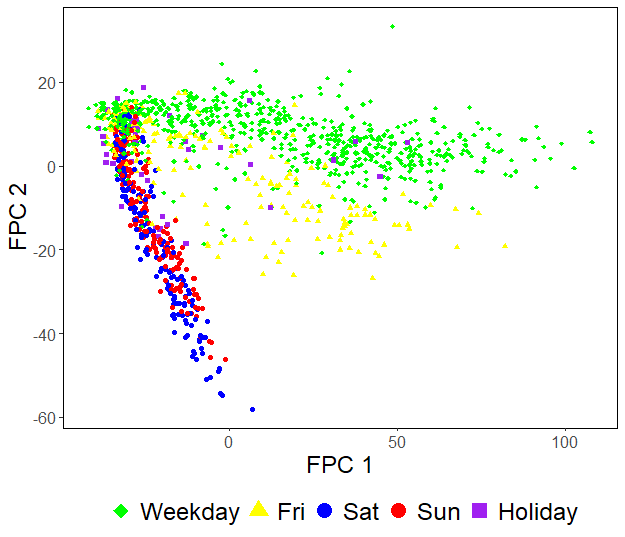}
\end{minipage}
\hfill
\begin{minipage}[c]{0.33\linewidth}
\includegraphics[width=\linewidth]{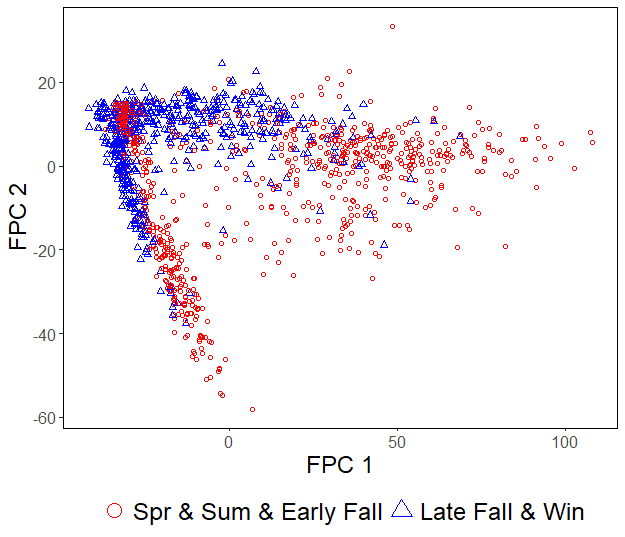}
\end{minipage}%
\hfill
\begin{minipage}[c]{0.33\linewidth}
\includegraphics[width=\linewidth]{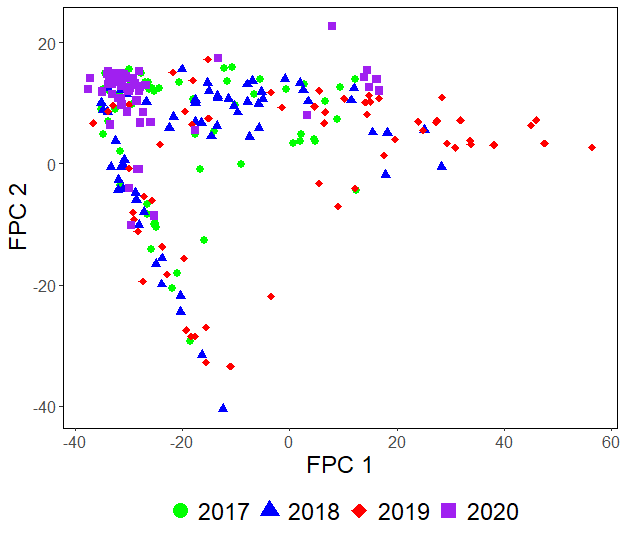}
\end{minipage}%
\caption{Pairwise plots of the first two functional principal component scores, distinguished by day of the week (left plot), by season (middle plot), and by year (right plot).\label{fig::FPC score}}
\end{figure}

Analysing the functional principal component  scores of the daily Fr\'{e}chet median distance trajectories along the first and second eigenfunctions, Fig. \ref{fig::FPC score} reveals three key patterns in the daily Fr\'{e}chet median distance trajectories. First, we observe that weekdays and weekends form distinct clusters, which can be seen in the left plot of Fig. \ref{fig::FPC score}. Second, seasonal differences  influence bike-sharing patterns, as shown in the  middle plot of Fig. \ref{fig::FPC score}. In the middle plot of Fig. \ref{fig::FPC score}, we display first versus second principal component scores, separated by two broad seasonal groups. Spring, summer, and early fall (April to October) exhibit greater variability than the late fall and winter months (November to March). This can be explained by the higher number of trips in the warmer months, which leads to greater variability, as shown in Figure \ref{fig::case study trip number} in the supplementary material. Finally, a similar pattern is observed in March and April from 2017 to 2020, with larger trip numbers corresponding to increased variability. In the right plot of Fig. \ref{fig::FPC score}, we display second versus first functional principal component scores, differentiated according to March and April in different years. From 2017 to 2020, the number of trips become larger as well as the variability in March and April. However, the smaller variability can be found in March and April in 2020. This is caused by the impact of COVID-19. This impact is notable, demonstrating a significant reduction in trips compared to previous years, which can be found in Fig. \ref{fig::case study trip number} in the supplementary material.

\begin{figure}
\begin{minipage}[c]{0.3\linewidth}
\includegraphics[width=\linewidth]{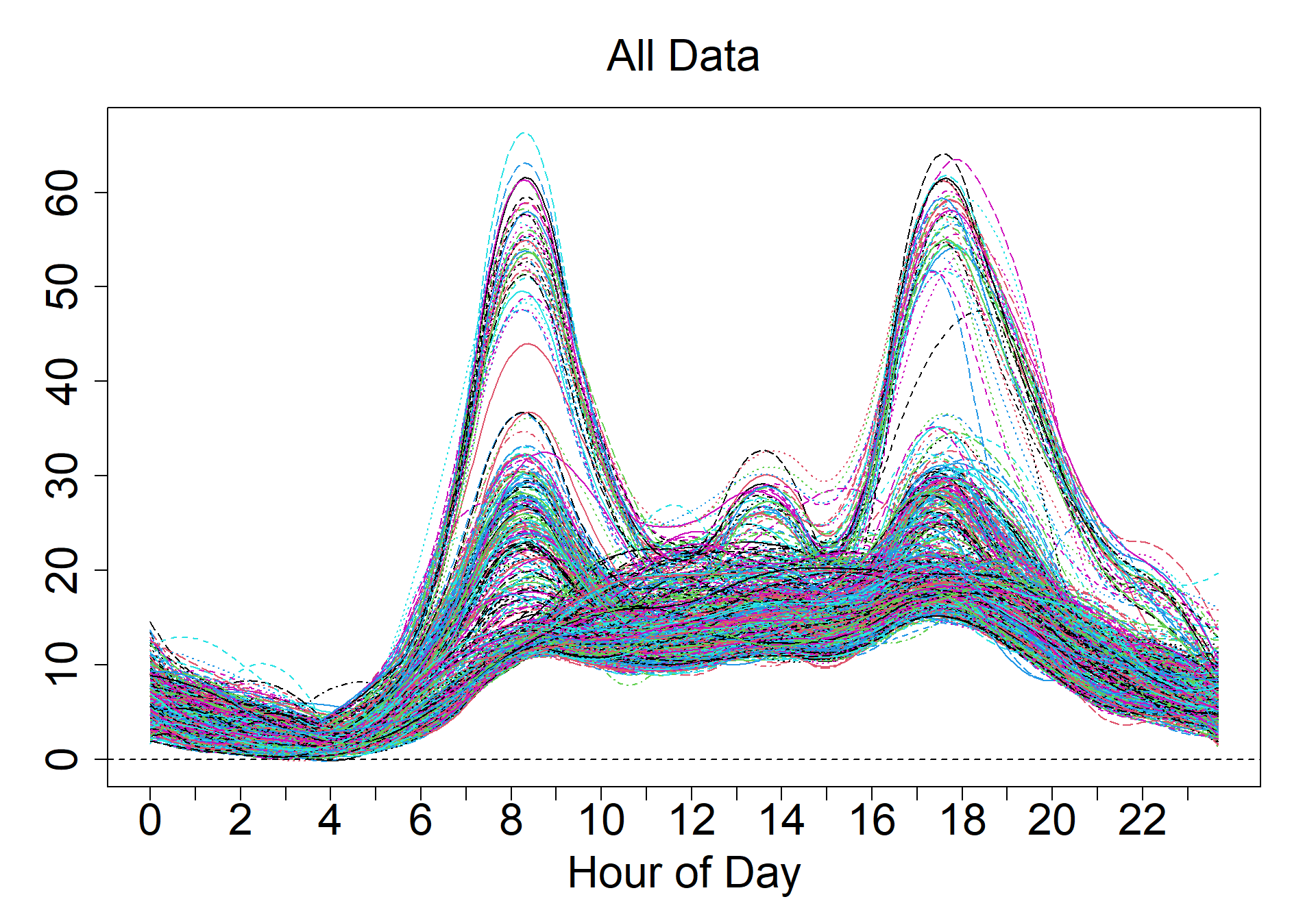}
\end{minipage}
\hfill
\begin{minipage}[c]{0.3\linewidth}
\includegraphics[width=\linewidth]{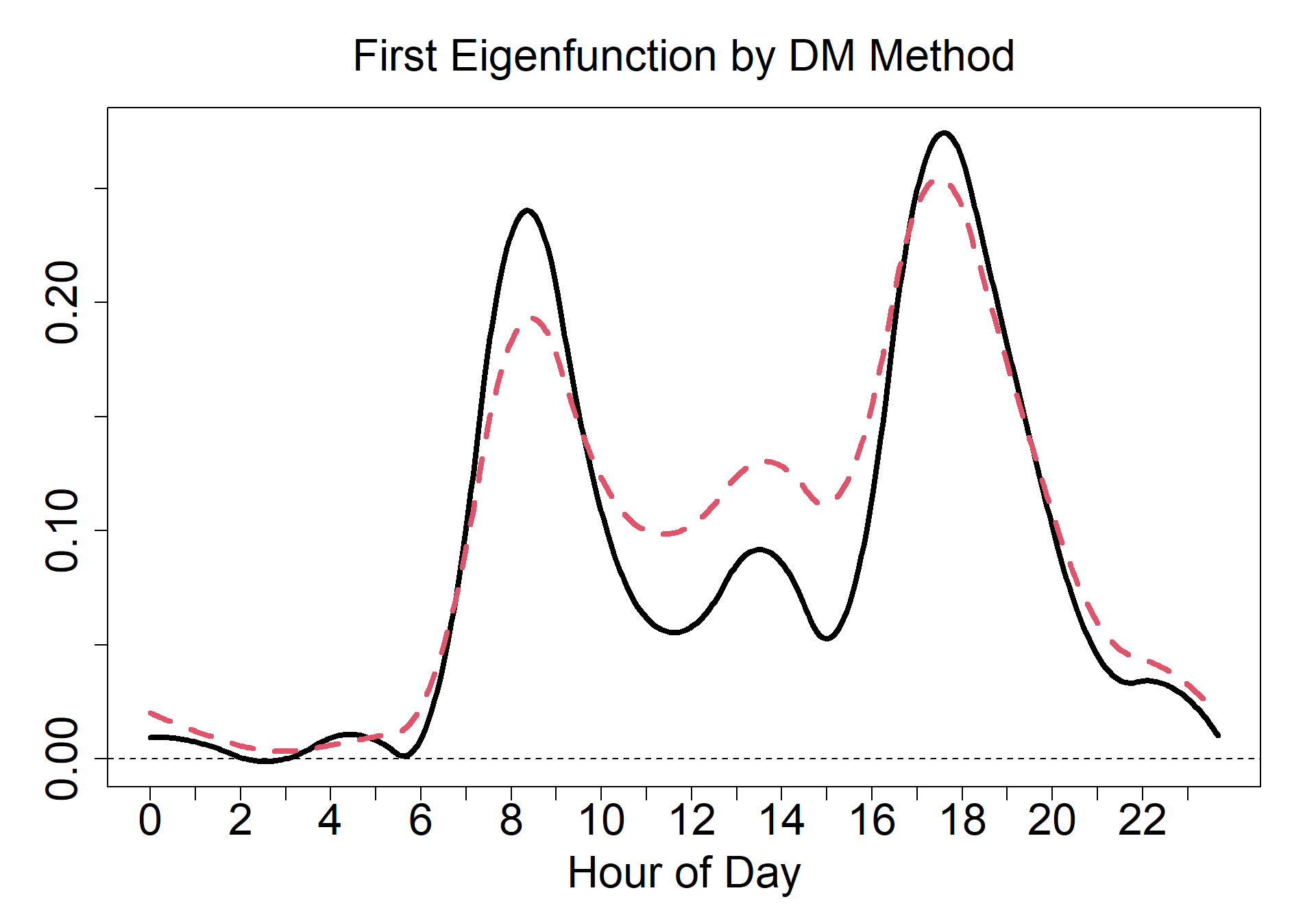}
\end{minipage}
\hfill
\begin{minipage}[c]{0.3\linewidth}
\includegraphics[width=\linewidth]{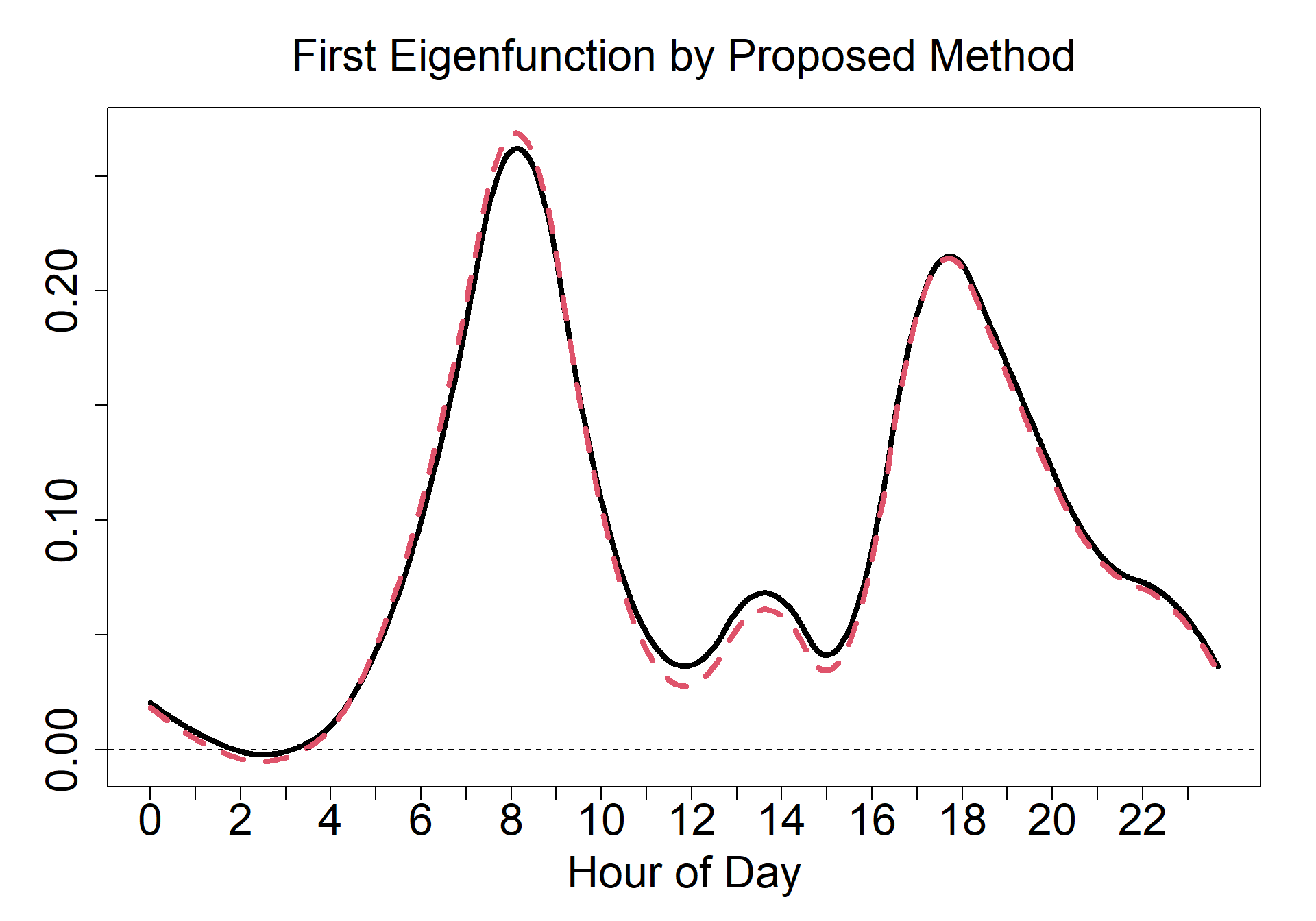}
\end{minipage}%
\vfill
\begin{minipage}[c]{0.3\linewidth}
\includegraphics[width=\linewidth]{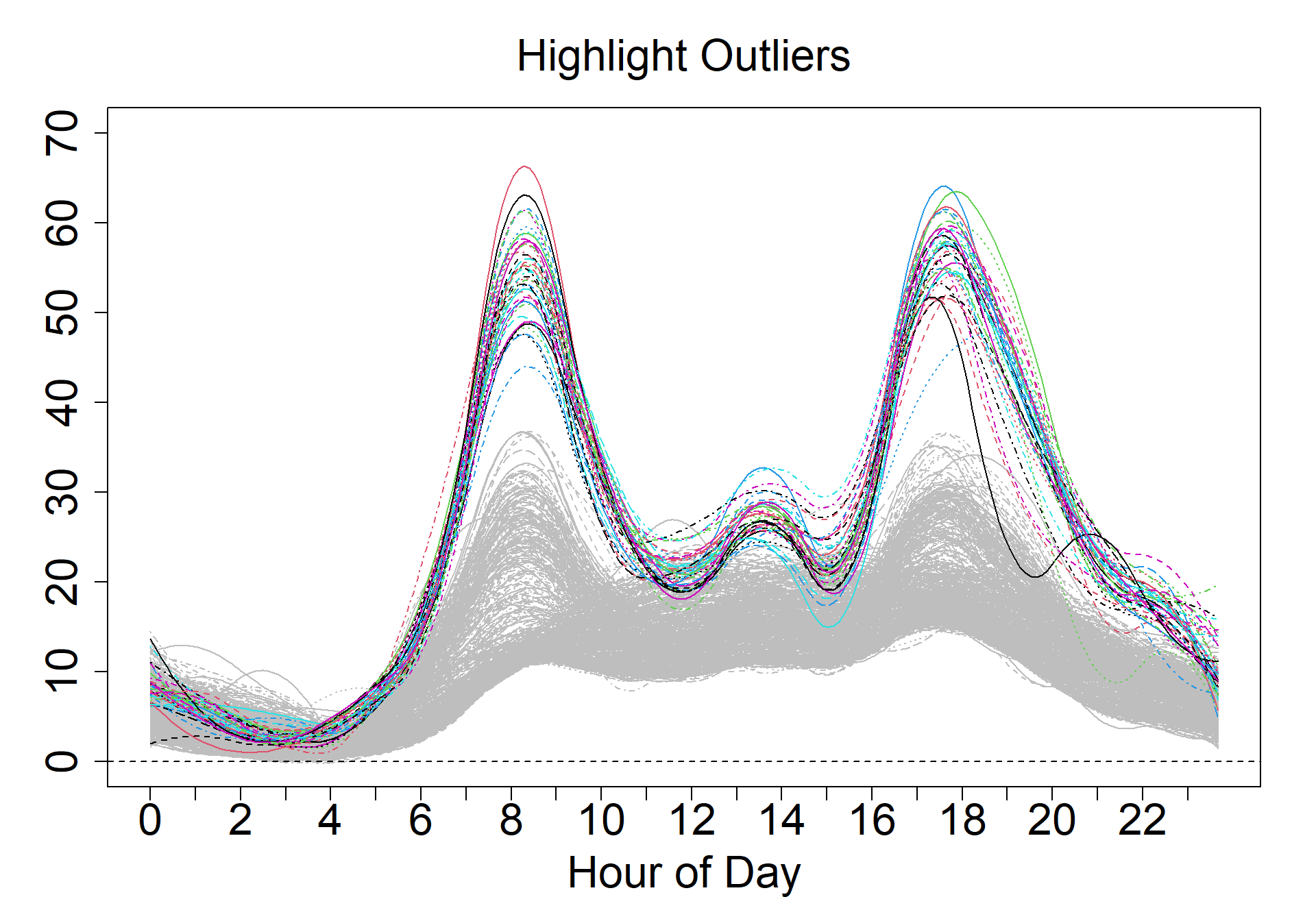}
\end{minipage}
\hfill
\begin{minipage}[c]{0.3\linewidth}
\includegraphics[width=\linewidth]{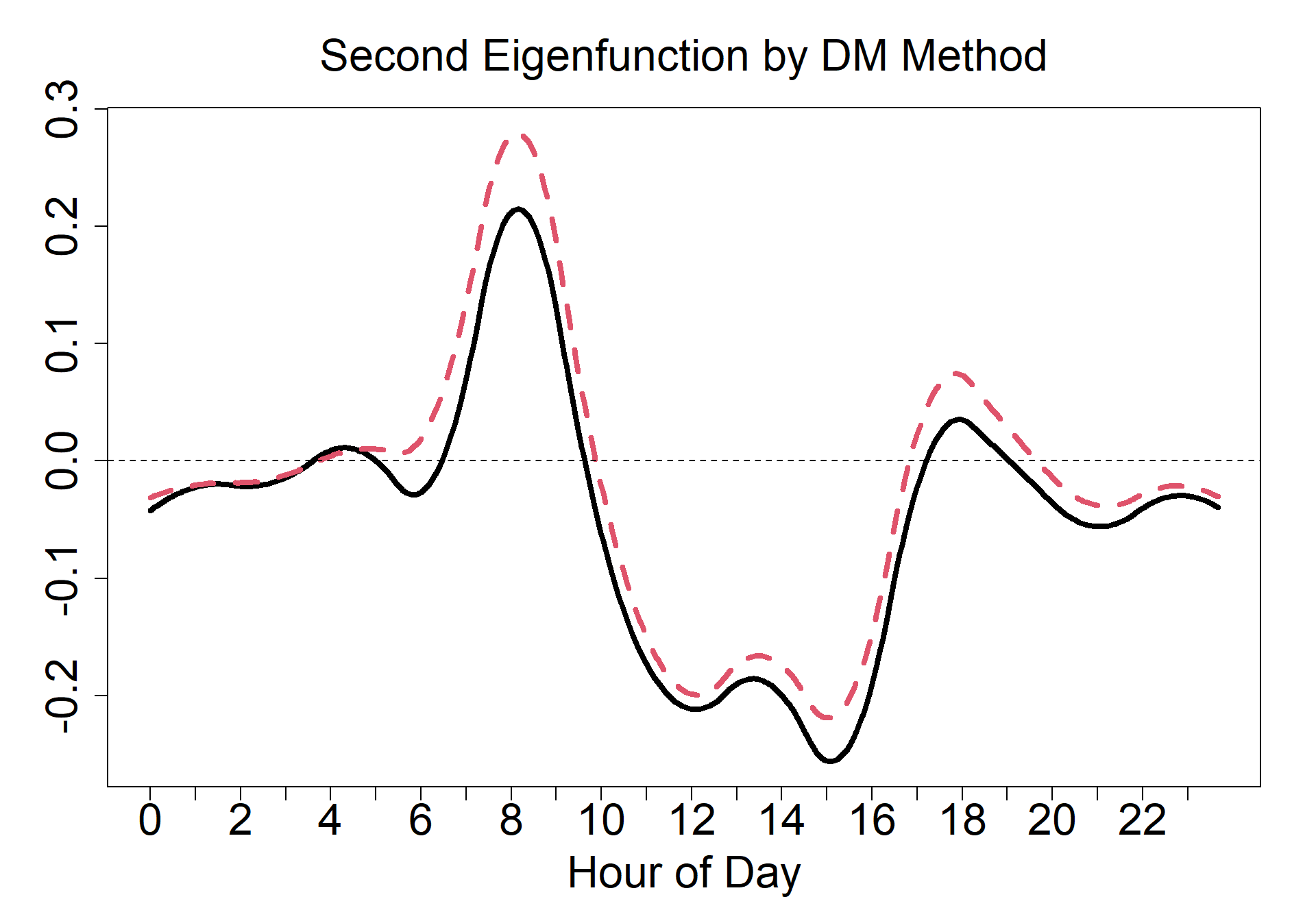}
\end{minipage}
\hfill
\begin{minipage}[c]{0.3\linewidth}
\includegraphics[width=\linewidth]{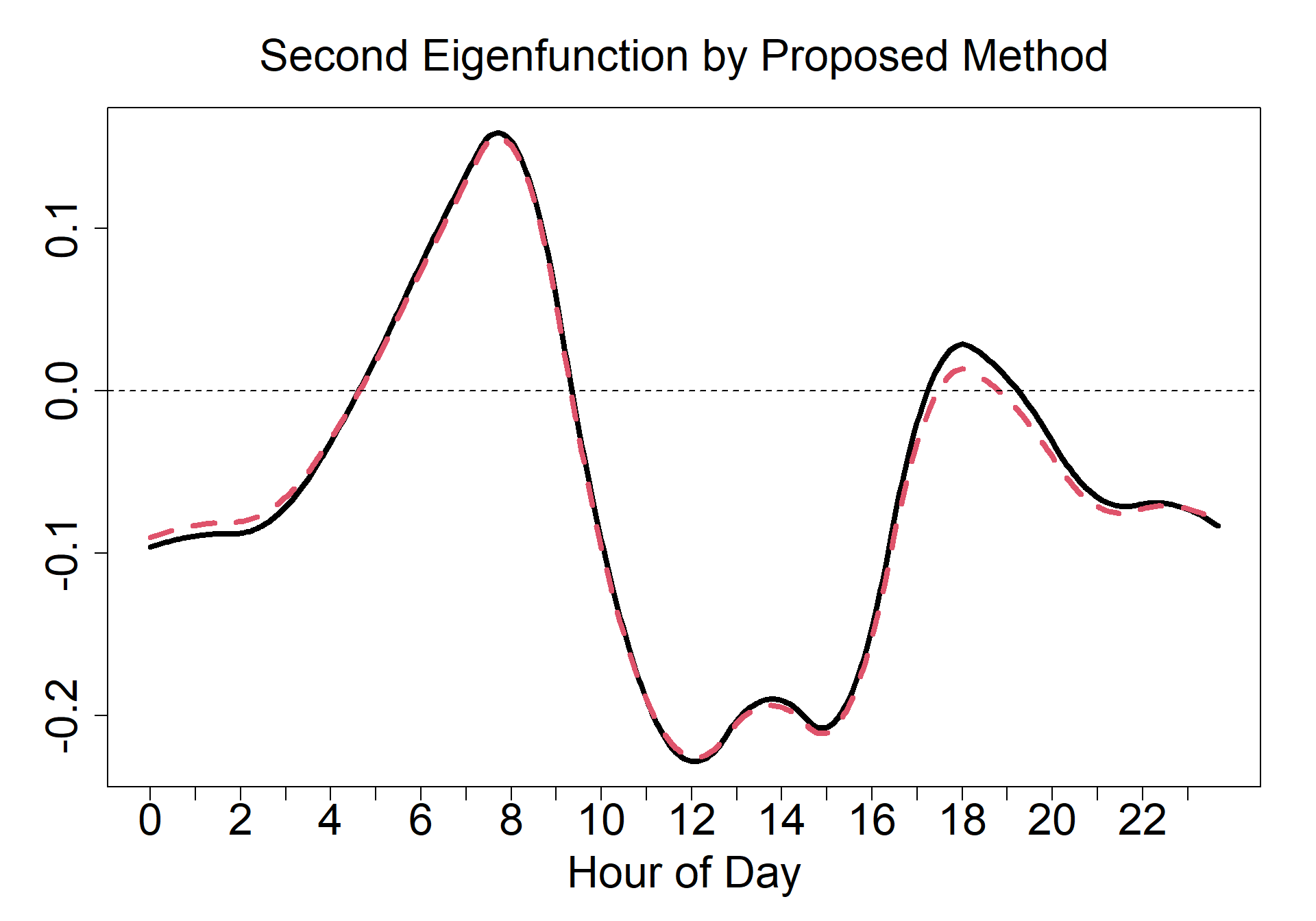}
\end{minipage}%
\caption{Estimated eigenfunctions of the  proposed autocovariance and the autocovariance in Dubey and M{\"u}ller's method, abbreviated as DM, with outliers included versus outliers removed. The solid black lines represent the estimated eigenfunctions with data including outliers, and the red dashed lines represent those with data excluding outliers. \label{fig::robust compare}}
\end{figure}

Finally, we investigate the robustness properties of Dubey and M{\"u}ller's method and our proposed method by performing analysis with and without outliers. For illustrative purposes, we focus on analysing a subset of 558 from the 1273 daily Fr\'{e}chet median distance trajectories, for easier visualization in demonstrating the influence of potential outliers on eigenfunction estimations. Figure \ref{fig::robust compare} displays the first two estimated eigenfunctions by our proposed autocovariance and the autocovariance in Dubey and M{\"u}ller's method. To compare the robustness of these two methods, the selection of the subset of daily Fr\'{e}chet median distance trajectories is based on their Frobenius norm, as the Fr\'{e}chet median distance trajectories convey the overall daily departure from the centre of the space of sample Laplacian matrices. The top left figure in Fig. \ref{fig::robust compare} shows the general shape of the spline smoothed 588 data, and the bottom left figure highlights 39 identified outliers. By comparing eigenfunctions estimated with and without outliers, our method can be seen to be robust, as the eigenfunctions overlap in Fig. \ref{fig::robust compare}. However, the estimated eigenfunctions by Dubey and M{\"u}ller's method are clearly more sensitive to the presence of outliers. Moreover, we can use the pairwise plots of the first two functional principal scores scores to detect outliers. Figure \ref{fig::case study outlier detect} in the supplementary material illustrates the successful identification of outliers using our proposed method through pairwise plots of the first two functional principal component  scores.

\section{Simulation Study}\label{sect::simulation study}

We illustrate our method by the simulation of samples of time-varying networks with 20 nodes in this section while simulations of samples coming from a metric space with non-negative curvature can be found in \S \ref{sect::sphere simulation} in the supplementary material. To facilitate comparison, we use a similar network generation technique to that in \cite{dubey2021modeling} with three different community structures. Details of data generation can be found in the supplementary material. We let the community memberships of the nodes stay fixed in time, while the edge connectivity strengths between the communities change with time. We first generate the network adjacency matrices $A_i(t)$ by (\ref{eq::adj matrix generate}) in the supplementary material and the trajectories are represented as graph Laplacians $X_i(t)=D_i(t)-A_i(t)$, where $D_i(t)$ is a diagonal matrix whose diagonal elements are equal to the sum of the corresponding row elements in $A_i(t)$. We generate 100 samples for each group and we have 300 samples in total. Adopting the Frobenius norm in the space of graph Laplacians, the Fr\'{e}chet median network at time $t$ is obtained via the extrinsic method whose details are given in the supplementary material, and we obtain the Frobenius distance trajectories of the individual subjects from the Fr\'{e}chet median trajectory. We then carry out robust functional principal component analysis of the generated distance trajectories.

\begin{figure}
\begin{minipage}[c]{0.3\linewidth}
\includegraphics[width=\linewidth]{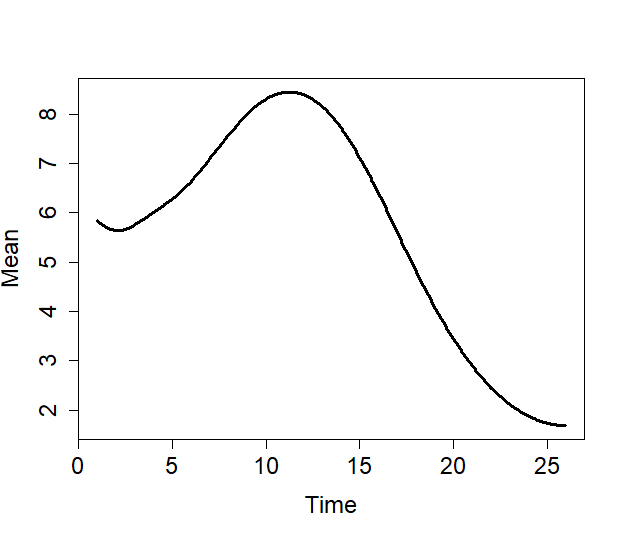}
\end{minipage}
\hfill
\begin{minipage}[c]{0.3\linewidth}
\includegraphics[width=\linewidth]{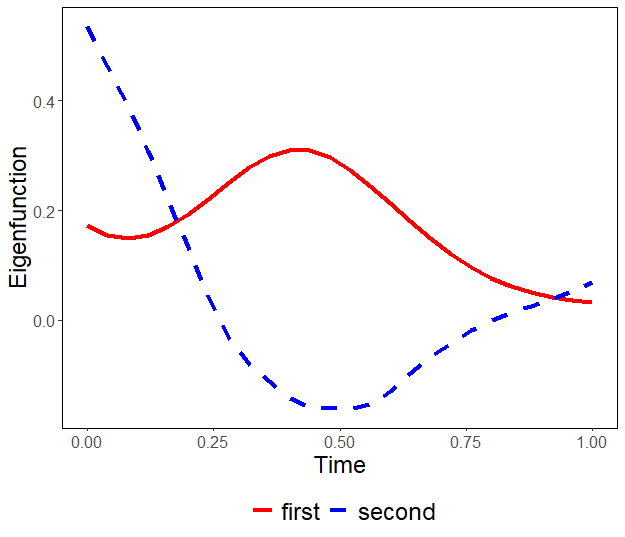}
\end{minipage}
\hfill
\begin{minipage}[c]{0.3\linewidth}
\includegraphics[width=\linewidth]{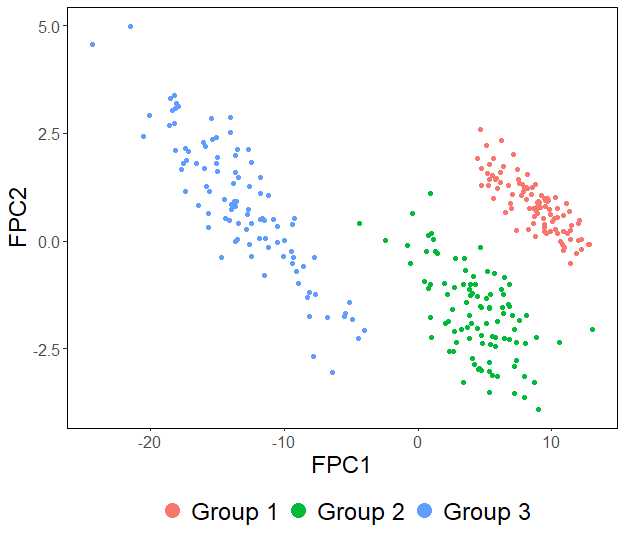}
\end{minipage}%
\caption{Sample mean function (left plot) and eigenfunctions (middle plot) and the scatter plot of first two functional principal component (right plot) obtained from the robust functional component analysis of the distance trajectories for networks.\label{fig::simulation sample mean eigenfunction biplot}}
\end{figure}

The results of robust functional principal component analysis are shown in Fig. \ref{fig::simulation sample mean eigenfunction biplot}. The proposed method is seen to perform well in recovering the groups in the scatter plot of the second versus first functional principal component. Groups 1 and 2 are found to have closer cluster centres than Groups 1 and 3. A detailed explanation may be found in the supplementary material.

To examine the finite-sample robustness of the proposed method, we also examine situations where a portion of the simulated data are contaminated by outliers. The percentages of outliers considered in our simulation are $\{0.05,0.1,\ldots,0.4\}$. We considered several types of contamination  but report only the worst-case scenario. Outliers are generated through adding a shift of 0.5 to the time varying connectivity weights and scaling the weights value by five times. In each settings above, 200 simulation replications were conducted.

\begin{figure}
\centering
\includegraphics[scale=0.34]{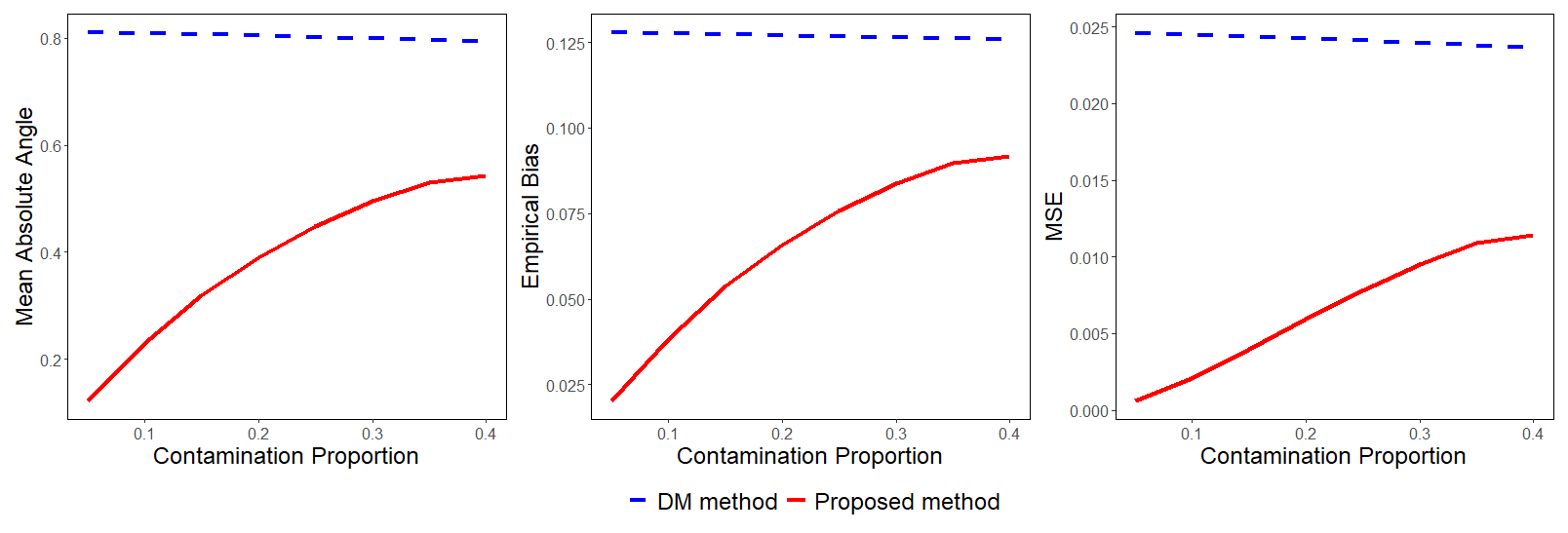}

\caption{Simulated data with outliers generated by extreme values with a shift in time varying connectivity weights. Mean absolute angle (left), empirical bias (middle) and Mean Squared Error (right) of the leading eigenfunction. DM method stands for Dubey and M{\"u}ller's method. \label{fig::simulation robust result}}
\end{figure}

We consider Dubey and M{\"u}ller's method proposed by \cite{dubey2021modeling} for comparison. The performance of the eigenfunction estimation is measured by the mean absolute angle, the empirical bias and the mean integrated square error. The mean absolute angle of the $k$-th eigenfunction is given by $\textrm{MEA}=\frac{1}{200}\sum_{r=1}^{200} \arccos (|\langle \hat{\phi}_k^{(r)},\bar{\phi}_k \rangle |)$ where $\hat{\phi}_k^{(r)}$ is the estimate of the $k$-th eigenfunction at the $r$-th replicate and $\bar{\phi}_k$ is the Monte Carlo approximation of the unknown population $k$-th eigenfunction under the uncontaminated setting. Figure \ref{fig::simulation robust result} shows simulation results on the estimation of the leading eigenfunction where our proposed method outperforms Dubey and M{\"u}ller's method. Additionally, as the contamination proportion gets closer to the breakdown point 0.4, all of the three measures of the leading eigenfunction increase slower as expected. Our proposed method outperforms Dubey and M{\"u}ller's method as their method starts to break down when the contamination ratio equal to 0.05. The reason for the failure of Dubey and M{\"u}ller's method can be found in Fig. \ref{fig::sim data with unimodal outliers} in the supplementary material. In Dubey and M{\"u}ller's method, the Fr\'{e}chet variance trajectories are the deviations from the Fr\'{e}chet mean trajectory and they are sensitive to the outliers. Figure \ref{fig::sim data with unimodal outliers} in the supplementary material shows that, due to the existence of outliers, the Fr\'{e}chet mean trajectory shifts to the left side, resulting in a shift of the Fr\'{e}chet variance trajectories. As there exist a large change in the shapes of Fr\'{e}chet variance trajectories after the contamination, it is not surprising that the Dubey and M{\"u}ller's method will break down with a small number of outliers.

It is worth noting that the type of outlying observation considered above relates almost exclusively to scale of network sizes. Another setting for outlying time-varying observations is studied in \S \ref{subsect::bimodal simulation} in the supplementary material, where we generate outlying time-varying networks with bimodal weighting functions in a sample of networks with unimodal weighting functions. Moreover, in\S \ref{subsect::zero weighting simulation} in the supplementary material, we study the case where the outliers have some zero weighting functions, allowing for changes in the connectivity structure .

We also provide numerical comparisons between the proposed method and the one using the pairwise spatial sign operator proposed in \cite{wang2023robust} in \S \ref{subsect::concentrated simulation} in the supplementary material. Note that combining the pairwise spatial sign operator proposed in \cite{wang2023robust} with the Fr\'{e}chet median distance trajectories results in the proposed autocovariance operator (\ref{swPCA cov}) with the cutoff point $Q$ being zero. Our proposed method is expected to outperform the method of \cite{wang2023robust} in cases when the time-varying networks have a single highly concentrated component. This is due to the fact that, when $Q=0$,  some ratios in (\ref{swPCA cov}) of the form  $\{V(s)-\widetilde{V}(s)\}\{V(t)-\widetilde{V}(t)\}/\|V(\cdot)-\widetilde{V}(\cdot)\|^2$ involve extremely small values in both the numerator and denominator, which can lead to numerical rounding errors in Wang's approach. Simulation results in the supplementary material support the above arguments.

\section{Discussion}
We present a framework for analysing time-varying random object data in the presence of outlying objects, where the random objects take values in a general metric space. Our approach introduces a generalized notion of the median function within the object space. A key component of this framework is the development of a novel robust functional principal component analysis autocovariance operator. This operator is specifically designed to handle the distance functions of subject-specific curves from the median function. The proposed robust estimator of the autocovariance operator is tailored to deal with the non-negative and dependence properties of the subject-specific curves from the median function.

Our numerical studies on network Laplacian matrix data show that our proposed robust functional principal component analysis can be used to detect clusters and outliers. Additionally, as we demonstrate in simulation studies and real data analysis, our proposed method outperforms the existing methodology for time-varying random object data (\citealt{dubey2021modeling}) with respect to robustness, but typically gives comparable performance when outlying objects are not present in the sample.

\section*{Supplementary material}
\label{SM}
The supplementary material consists of overview of the theoretical derivations, discussion on the conditions used in deriving the uniform convergence of empirical barycenters, technical lemmas, proofs of theorems, propositions and corollaries, additional simulation studies on network Laplacians and spherical data, additional figures for the case study, and numerical studies on the potential information loss after the distance-based transformation on univariate and multivariate functional data.

\vspace*{-10pt}

\appendix

\appendixone
\section*{Appendix}
This Appendix includes eight useful assumptions, the discussion on the choice of the metric and some examples of metric spaces which satisfy the assumptions. The assumptions are discussed below.

\begin{assumption}
\label{condition c0}
For each $t\in[0,1]$, there exists a $\eta>0$ such that $\sup_{\omega\in\bar{\Omega}}E\left( \int_0^1 |d(\omega,X(t))|^{2+\eta}dt\right)<\infty$ for any compact subset $\bar{\Omega}$ of $\Omega$.
\end{assumption}

\begin{assumption}
\label{condition c1}
For each $t\in[0,1]$, the pointwise Fr\'{e}chet median $\mu_{\rm GM}(t)$ given in (\ref{def::frechet median}) exists and is unique, and
  \[
    \inf_{t\in[0,1]}\inf_{\omega\in\Omega:d(\omega,\mu_{\rm GM}(t))>\gamma} E  [d(\omega,X(t))]- E  [d(\mu_{\rm GM}(t),X(t))]>0
  \]
  for any $\gamma>0$. 
\end{assumption}

\begin{assumption}
\label{condition c1.5}
$(\Omega,d)$ is a convex metric space, i.e., for every $x,y\in\Omega$, there exists $z\in\Omega$ such that $d(x,z)+d(z,y)=d(x,y)$. Define $\bar{M}_n(\omega;t)=\frac{1}{n}\sum_{i=1}^n d(\omega,X_i(t))$ and suppose there exists $\widetilde{\mu}(t)$ such that $d(\mu_{\rm GM}(t),\widetilde{\mu}(t))+d(\widetilde{\mu}(t),\hat{\mu}_{\rm GM}(t))=d(\mu_{\rm GM}(t),\hat{\mu}_{\rm GM}(t))$ and $\bar{M}_n(\widetilde{\mu}(t);t) \leq \zeta(t) \bar{M}_n(\hat{\mu}_{\rm GM}(t);t) + (1-\zeta(t)) \bar{M}_n(\mu_{\rm GM}(t);t)$  holds for all $t$ where $\zeta(t)=d(\mu_{\rm GM}(t),\widetilde{\mu}(t))/d(\mu_{\rm GM}(t),\hat{\mu}_{\rm GM}(t))$.
\end{assumption}

\begin{assumption}
\label{condition c2}
There exists $\rho>0$, $D>0$ and $\beta>1$ such that
  \[
    \inf_{t\in[0,1]}\inf_{\omega\in\Omega:d(\omega,\mu_{\rm GM}(t))<\rho}\left\{   E  [d(X(t),\omega)]- E  [d(X(t),\mu_{\rm GM}(t))]-Dd^\beta(\omega,\mu_{\rm GM}(t)) \right\}\geq0,
  \]
\end{assumption}

\begin{assumption}
\label{condition c3}
For some $0<\alpha\leq 1$, the random function $X(\cdot)$ is defined on $[0,1]$ and takes values in $\Omega$ and suppose to be $\alpha$-H\"{o}lder continuous. Denote the space of all such functions as $\Omega^{[0,1]}$. That is, for non-negative $G:\Omega^{[0,1]}\to\mathbb{R}^+$ with $ E  [G(X(\cdot))^2]<\infty$, it holds almost surely,
  \[
    d(X(s),X(t))\leq G(X(\cdot))|s-t|^\alpha.
  \]
\end{assumption}

\begin{assumption}
\label{condition c4}
For $I(\delta)=\int_0^1\sup_{t\in[0,1]}\sqrt{\log N(\varepsilon\delta,B_\delta(\mu_{\rm GM}(t)),d)}d\varepsilon$, it holds that $I(\delta)=O(1)$ as $\delta\to0$, where $B_\delta(\mu_{\rm GM}(t))=\{\omega\in\Omega:d(\omega,\mu_{\rm GM}(t))<\delta \}$ is the $\delta$-ball around $\mu_{\rm GM}(t)$ and $N(\gamma,B_\delta(\mu_{\rm GM}(t)),d)$ is the covering number.
  
\end{assumption}

\begin{assumption}
\label{condition c5}
For each $j\geq 1$, the eigenvalue $\lambda_j(C)$ has multiplicity one, i.e., it holds that $\delta_j>0$ where $\delta_j=\min_{1\leq l\leq j}(\lambda_l-\lambda_{l+1})$.
\end{assumption}

\begin{assumption}
\label{condition c6}
 $V(\cdot)$ is \textit{weakly functional coordinate symmetric}. That is, for any positive integers $d\leq d^\prime \leq \infty$ with $d<\infty$ and any orthonormal bases $\{\psi_1(\cdot),\ldots,\psi_d(\cdot)\}$ in $\mathcal{H}$, $(\langle V(\cdot)-\widetilde{V}(\cdot),\psi_1(\cdot)\rangle,\ldots,\langle V(\cdot)-\widetilde{V}(\cdot),\psi_d(\cdot)\rangle)^\top=\Psi Z_\psi$ in distribution, where $\Psi$ is a $d\times d^\prime$ matrix only depending on $\{\psi_1(\cdot),\ldots,\psi_d(\cdot)\}$ such that $\Psi\Psi^\top=I_d$, the $d$-dimensional identity matrix, and $Z_\psi\in\mathbb{R}^{d^\prime}$ is coordinatewise symmetric in the sense that $GZ_\psi=Z_\psi$ in distribution for any diagonal matrix $G$ with diagonal elements $G_{jj}\in\{-1,1\}$.

\end{assumption}

Assumption \ref{condition c0} is a moment condition that is slightly stronger than the standard assumption in functional data analysis that $\sup_{\omega\in\Omega}E\left( \int_0^1 (d(\omega,X(t)))^2dt\right)<\infty$ for all $t\in[0,1]$. Assumption \ref{condition c0} is necessary because we only require that $\Omega$ is totally bounded in a neighbourhood of the population Fr\'{e}chet median, whereas most existing works, such as  \cite{dubey2020functional,dubey2021modeling}, assume that $\Omega$ itself is totally bounded. A global total boundedness assumption is often too restrictive in robust analysis. Assumption \ref{condition c1} is an identifiability condition for the population Fr\'{e}chet median.  The sample verison belongs to the class of $M$-estimators (see Theorem 5.7 in \cite{vd98asym} for further discussion). Assumption \ref{condition c1.5} imposes a local convexity condition on the empirical cost function $\bar{M}_n(\omega;t)$ defined on the metric space in a  neighbourhood of the population Fr\'{e}chet median. If $\Omega$ is a convex subset of the Euclidean space with the induced Euclidean norm, this assumption aligns with the standard local convexity condition in M-estimator theory. Assumption \ref{condition c1.5} is a natural generalization of local convexity for functions on linear spaces to functions on convex metric spaces;  see \cite{abdelhakim2016convexity} and \cite{khamsi2011introduction} for discussion of the latter. Assumptions \ref{condition c0}-\ref{condition c1.5} guarantee the uniform convergence of the sample Fr\'{e}chet median trajectory to its population target, as they imply $\sup_{t\in[0,1]}d(\hat{\mu}_{\rm GM}(t),\mu_{\rm GM}(t))=o_p(1)$ with Assumption \ref{condition c0} ensuring that the dominated convergence theorem holds.  Importantly, we do not require the sample Fr\'{e}chet median belongs to a compact set. Furthermore, Assumption \ref{condition c1} is weaker than the assumptions in existing works such as \cite{dubey2020functional,dubey2021modeling} as we only assume the identifiability condition for the population Fr\'{e}chet median, which follows from the existence and uniqueness of the population Fr\'{e}chet median.

 As the sample Fr\'{e}chet median function is a special case of the cost functions in $M$-estimation, measurability issues can be addressed using outer probability measures for M-estimators, as discussed in \cite{vdVW96}. Assumption \ref{condition c2} is a standard requirement for M-estimators, which characterizes the local curvature of the cost function near its minimum. The curvature is characterized by $\beta$, which controls the convergence rate of the Fr\'{e}chet median. Assumption \ref{condition c3} is a mild smoothness assumption, commonly satisfied for certain values of $\alpha$ by many common Euclidean-valued random processes. Assumption \ref{condition c4} is a bound on the covering number, defined e.g. in \cite{vdVW96}, of the object metric space and is satisfied by several commonly encountered random objects. Assumption \ref{condition c5} is needed to show the asymptotic property of the estimated eigenfunctions. Notably, Assumptions \ref{condition c1}, \ref{condition c2}-\ref{condition c5} are consistent with those used in \cite{dubey2020functional,dubey2021modeling}. Object spaces that meet these conditions include graph Laplacians of networks with the Frobenius metric, univariate probability distributions with the 2-Wasserstein metric, and correlation matrices of fixed dimensions with the Frobenius metric (see \cite{petersen2019frechet}). Moreover, these metric spaces satisfy Assumption \ref{condition c2} with $\beta=2$ (see \citealt{dubey2021modeling}). Assumption \ref{condition c6} represents the weakest condition in the literature, to our knowledge, ensuring that the robust estimator of the autocovariance operator shares the same eigenfunctions as the regular covariance function. This condition, introduced by \cite{wang2023robust}, generalizes the functional coordinate symmetry condition from \cite{gervini2008robust}. \cite{wang2023robust} demonstrates that Assumption \ref{condition c6}, the weak functional coordinate symmetry condition, permits arbitrary marginal distributions.

The choice of a metric is crucial as it not only affects relevant notions of outlyingness but also affects the Fr\'{e}chet median distance trajectories. Some basic criteria for metric selection include (i) feasibility and ease of calculation implementation of the  Fr\'{e}chet median, and (ii) the metric space with the chosen metric should satisfy the required assumptions given in the Appendix, especially Assumption \ref{condition c1.5}. For example, the space of network Laplacian with the Frobenius norm satisfies both (i) and (ii) while the  Fr\'{e}chet median calculation is not currently feasible for the space of network Laplacian with the Procrustes size-and-shape distance (\citealt{severn2022manifold}). Below are examples of some metric spaces satisfying Assumption \ref{condition c1.5}.

\begin{example}[Networks]
For the space of graph Laplacians of networks with the Frobenius norm, the distance between two networks is equivalent to the distance between the half-vectorization of the corresponding networks with the Euclidean norm. By \cite{khamsi2011introduction}, one can see that it is the convex metric space. Moreover, the Euclidean norm is convex and thus the empirical cost function $\bar{M}_n(\omega;t)$ is also convex. Therefore, Assumption \ref{condition c1.5} holds.
\end{example}

\begin{example}[Spheres]
For observations coming from the Euclidean sphere $\Omega = \mathcal{S}^2$ in $\mathbb{R}^3$ equipped with the great-circle distance, i.e., for $p_1,p_2\in \mathcal{S}^2$, $d(p_1,p_2)={\rm arccos}(p_1^\top p_2)$. By \cite{khamsi2011introduction}, one can see that it is the convex metric space. When the observations are concentrated such that the maximum distance between any two observations is less than $\pi/2$, the empirical cost function $\bar{M}_n(\omega;t)$ is locally convex by \cite{fletcher2009geometric}.
\end{example}

\section*{Acknowledgement}
We thank an associate editor and two reviewers for constructive comments. The work of Andrew T. A. Wood was supported by Australian Research Council grant DP220102232.

\bibliographystyle{biometrika}

\end{document}